\documentclass[9pt,twocolumn,twoside]{opticajnl}
\journal{opticajournal} 

\setboolean{shortarticle}{true}

\usepackage{lineno}

\title{RF E-field enhanced sensing based on Rydberg-atom-based superheterodyne receiver}

\author[1,2]{Wenguang Yang}
\author[1,2]{Minyong Jing}
\author[1,2]{Hao Zhang}
\author[1,2,*]{Linjie Zhang}
\author[1,2]{Liantuan Xiao}
\author[1,2]{Suotang Jia}

\affil[1]{State Key Laboratory of Quantum Optics and Quantum Optics Devices, Institute of Laser Spectroscopy, Shanxi University, Taiyuan 030006, People’s Republic of China}
\affil[2]{Collaborative Innovation Center of Extreme Optics, Shanxi University, Taiyuan, Shanxi 030006, People’s Republic of China}
\affil[*]{zlj@sxu.edu.cn}

\begin{abstract}
We present enhanced sensing of radio frequency (RF) electric fields (E-fields) by the combined polarizability of Rydberg atoms and the optimized local oscillator (LO) fields of supergheterodyne receiving. Our modified theoretical model reveals the dependencies of sensitivity of E-field amplitude measurement on the polarizability of Rydberg states and the strength of the LO RF field. The enhanced sensitivity of megahertz(MHz) E-field are demonstrated at an optimal LO field for three different Rydberg states $\rm 43D_{5/2}$, $\rm 60S_{1/2}$, and $\rm 90S_{1/2}$. The sensitivity of 63 MHz for the $\rm 90S_{1/2}$ state reaches 0.96 $\mu \rm V/cm/\sqrt{Hz}$ that is about an order of magnitude higher than those already published. This result closely approaches the theoretical sensitivity limit of RF dipole antennas, and indicates the potential for breaking the limit in measuring sub-MHz E-fields. This atomic sensor based on Rydberg Stark effect with heterodyne technique is expected to boost an alternative solution to electric dipole antennas.\par

\end{abstract}

\setboolean{displaycopyright}{false} 

\begin{document}

\maketitle

\section{Introduction}
Rydberg atoms possess enormous electric dipole moments and polarizabilities, making them highly sensitive to external E-field. When combined with Electromagnetically Induced Transparency (EIT) technology, they can be used for precise measurements of RF E-field. The new appeal of using Rydberg atoms as a medium for sensing RF E-field lies in the fact that it not only offers a large frequency bandwidth\cite{holloway2014broadband,gordon2014millimeter,jau2020vapor} but also provides an ultra high sensitivity \cite{jing2020atomic, Borowka2023NP} and further a wide dynamic range\cite{anderson2014two,jing2020atomic} for field strength sensing. Furthermore, Rydberg-EIT experiments can be conducted in room-temperature vapor cells, making such measurement systems more compact and portable compared to cold atomic systems. Many research groups have initiated research efforts with the aim of establishing RF E-field measurement standards based on Rydberg atoms\cite{HollowayImagingReview,HollowayPowerStandard,HollowayRaithelLinearity}. These studies involve investigations in terms of the amplitude\cite{jing2020atomic,gordon2010quantum,sedlacek2012microwave}, polarization\cite{sedlacek2013atom,simons2019embedding}, and phase\cite{jing2020atomic,simons2019rydberg} of RF E-field.\par
The MHz E-field has longer wavelengths and extended propagation distances than the GHz E-field, which is particularly important for international and regional broadcasting, as well as aviation-air-to-ground communication. According to Chu limit theorem\cite{chu1948physical}, in the field of communication, there is a trade-off between the bandwidth and radiation efficiency of transmitting antennas. For receiving antennas, the focus is on sensitivity and bandwidth. Additionally, the maximum bandwidth achievable by a small antenna is directly proportional to the size of the antenna. The channel capacity is limited\cite{cox2018quantum} as the size of the dipole antenna is much smaller than the wavelength of the electromagnetic wave. In contrast, Rydberg atomic sensors are not limited by size. The vapor cells used are generally only a few centimeters long. The data capacity of the sensor far exceeds that of traditional antennas of the same size. Therefore, it holds great promises of atomic sensor for ultra high sensitive detection and communication in MHz band.\par
The process of measuring RF E-field based on the Stark effect involves preparing atoms in a superposition state sensitive to the E-field through optical pumping. The interaction between the external field and the superposition state leads to a frequency shift of energy levels as depicted in fig.\ref{fig:Schematic of setup}(a). Subsequently, an optical readout method is used to deduce the characteristics of the RF E-field\cite{liu2022highly,anderson2017continuous,jau2020vapor}. However, it is difficult to measure the extremely weak field as the frequency shift is trivial since the energy level shift and perturbation of the Rydberg atomic population caused by the weak E-field is very small. In the experimental implementation, it is feasible to optimize sensitivity by introducing an additional LO field, denoted as a conduction parameter $\gamma$\cite{degen2017quantum}, to maximize the response to the signal field. In this context, our comprehensive modified model establishes a relationship between the response to the external field, specifically the EIT transmission spectra to the intensity of the signal field, allowing us to determine the optimal LO field strength. Furthermore, we conducted measurements of the RF E-field under three principal quantum numbers of Rydberg states and revealed that the sensitivity is determined by the joint effect of polarizability and strength of LO field.

\section{Theory}

\begin{figure}[ht]
\centering
\fbox{\includegraphics[width=\linewidth]{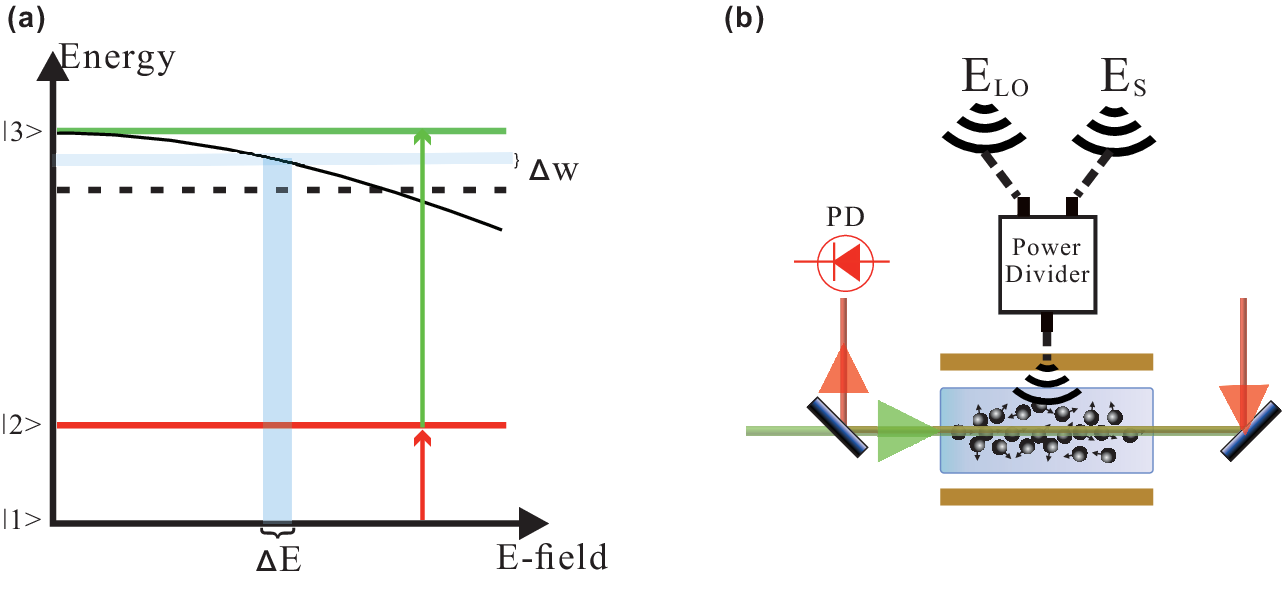}}
\caption{(a) Three-level ladder-type Cs energy diagram. An probe laser drives the transition $|1\rangle \rightarrow |2\rangle$ and coupling laser drives the transition $|2\rangle \rightarrow |3\rangle$. The change in E-field $\Delta \mathrm{E}$ corresponds to the change in energy level $\Delta \rm w$.(b) Schematic of experiment apparatus. The LO field $\mathrm{E}_{\mathrm{LO}}$ and the signal field $\mathrm{E}_{\mathrm{S}}$ are combined through a power divider and then fed into parallel electrode plates.The probe light enters the photodetector after passing through the vapor cell.}
\label{fig:Schematic of setup}
\end{figure}

Fig.\ref{fig:Schematic of setup} shows the energy-level structure and schematic layout of experiment apparatus. In our experiments, we focus on EIT in the 3-level-ladder system in which the probe laser (852 nm) is resonant with transition $6S_{1/2}(F=4) |1\rangle \rightarrow  6P_{3/2}(F^{'}=5) |2\rangle$ and the coupling laser is scanned through the transition $6P_{3/2}(F^{'}=5)|2\rangle\rightarrow43D_{5/2}|3\rangle$ of Cesium(Cs) atoms, the energy levels shift in the presence of RF E-field. The LO E-field and signal field are combined through a power divider and then fed into the space between a pair of parallel electrode plates as illustrated in fig.\ref{fig:Schematic of setup}(b). The EIT transmission spectra obtained by the photodetector is sent to an oscilloscope or the heterodyne signal is sent to a spectrum analyzer for analysis.\par
\par
The Stark shift of the Rydberg state depends on the polarizability $\alpha$ of the atoms:\par
\begin{equation}
\mathrm{w} = \frac{1}{2}\alpha\langle \mathrm{E}\rangle^{2}
\label{eq:Stark shift}
\end{equation}\par
In the heterodyne detection configuration, the LO RF E-field is represented as $\mathrm{E}_{\mathrm{LO}} \cos \left(f t+\varphi_{\mathrm{LO}}\right)$ and the signal RF E-field as $\mathrm{E}_{\mathrm{S}} \cos [(f+\Delta f) t]$.\par
Taking the time average of the above values, the difference frequency  $\Delta f$ is less than the instantaneous bandwidth of the atomic sensor. The electric-field-dependent Stark shift $\delta_c$ can be
written as:
\begin{equation}
    \delta_c=\bar{\delta}_c+\delta_c(\mathrm{t})=\frac{1}{2} \alpha\left[\frac{\mathrm{E}_{\mathrm{S}}^2}{2}+\frac{\mathrm{E}_{\mathrm{LO}}^2}{2}\right]+\frac{1}{2} \alpha \mathrm{E}_{\mathrm{LO}} \mathrm{E}_{\mathrm{S}}  \cos \left(\Delta f \mathrm{t}+\varphi_{\mathrm{LO}}\right)
\label{eq:Stark shift1}
\end{equation}\par
From eq. \ref{eq:Stark shift1}, it can be seen that the Stark shift includes both the DC (direct current) component and the AC (alternating current) component. Moreover, the product of the polarizability and the LO field strength in the AC component provides gain $\frac{1}{2} \alpha \mathrm{E}_{\mathrm{LO}}$ to the signal field.\par
Since $\mathrm{E}_{\mathrm{S}}$ is much smaller than $\mathrm{E}_{\mathrm{LO}}$, the DC component $\mathrm{E}_{\mathrm{LO}}$ serves as a parameter to adjust the slope of Stark shift. As demonstrated in fig. \ref{fig:stark map} and eq. \ref{eq:Stark shift1}, the slope of the Stark shift increases when LO field is stronger. More importantly, the signal we directly measure is the response of Rydberg EIT transmission spectra to the E-field. Therefore, we will practice combining the E-field and the EIT window. In order to understand the sensitivity more intuitively, without considering Doppler broadening and optically thin samples, we have derived an analytical solution based on the measurement scheme of RF heterodyne. First, we define the laser power after passing through the atoms as $\rm P_{out}$:\par
\begin{equation}
    \mathrm{P_{out}} = \mathrm{T \cdot P_{in}}
\label{eq:Pout}
\end{equation}\par
Where $\mathrm{P_{in}}$ is the probe laser power incident on the vapor cell, T is the transmission coefficient. The ultimate sensitivity of our measurement scheme is reached when the change of $\mathrm{P_{out}}$ equals the noise of probe laser $P_{noise}$.\par
\begin{equation}
    \Delta \mathrm{P}_{\text {out }}=\mathrm{P}_{\text {noise }}=\kappa \cdot e^{-l \operatorname{Im}\left[\rho_{21}\left(\Delta \delta_{\mathrm{c}}\right)\right]} \cdot \mathrm{P}_{\text {in }}
\label{eq:Delta-Pout}
\end{equation}\par
Here $\kappa=\frac{d}{d \delta_{\mathrm{c}}} e^{-l \operatorname{Im}\left[\rho_{21}\left(\delta_{\mathrm{c}}\right)\right]}$ and $l$ is cell length.
By solving  the master equation using the semi-classical density matrix approach\cite{fleischhauer2005electromagnetically}, one can obtain the density matrix element $\rho_{21}$ that describes the atomic coherence between the ground state and the first excited state.\par
\begin{equation}
    \rho_{\mathrm{ab}}=-\frac{i \Omega_{p r} / 2}{\gamma_{a b}-i \Delta_p+\frac{\Omega_c^2 / 4}{\gamma_{31}-i\left(\Delta_p+\delta_c\right)}}
\label{eq:rho21}
\end{equation}\par
Substitute eq. \ref{eq:Stark shift1} and \ref{eq:rho21} to eq. \ref{eq:Delta-Pout} to obtain the minimum detectable E-field(equal to the sensitivity when bandwidth is 1 Hz) as a function of amplitude of the signal field and the LO E-field, and also signal-to-noise ratio(SNR) is directly proportional to the gain controlled by the LO E-field:  \par
\begin{equation}
    \Delta \mathrm{E}_{\min } \propto \frac{1}{\frac{d \operatorname{Im}\left[\rho_{\mathrm{ab}}\left(\mathrm{E}_{\mathrm{LO}}, \mathrm{E}_{\mathrm{S}}\right)\right]}{d \mathrm{E}_{\mathrm{s}}}} \propto \frac{1}{\mathrm{SNR}}
\label{eq:sensitivity}
\end{equation}\par
This model will demonstrate the measurement sensitivity depend on the presence of an optimal operating point for the LO E-field, as illustrated in the following sections.\par
\section{Results and Discussion}

\begin{figure}[ht]
\centering
\fbox{\includegraphics[width=0.7\linewidth]{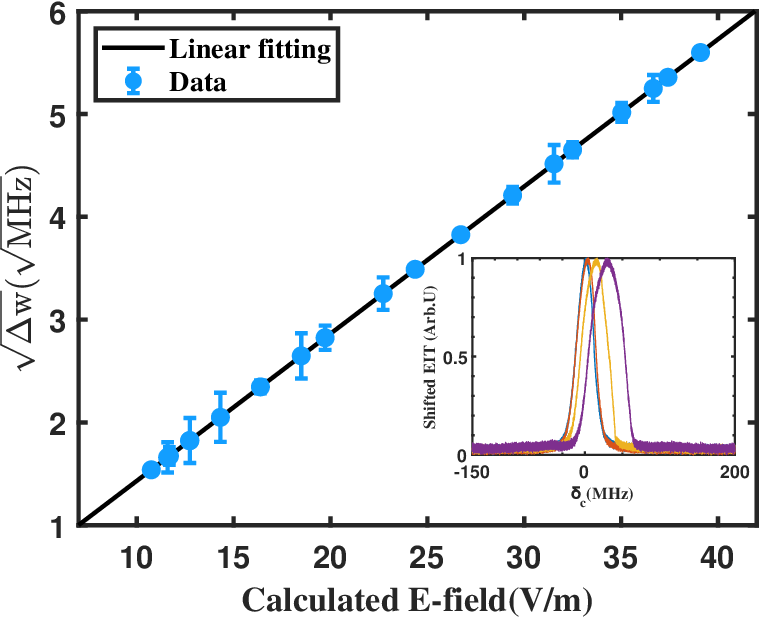}}
\caption{The square root of Stark shift Vs. the calculated E-field. The illustration depicts the Rydberg-EIT spectra versus detuning of coupling laser at different E-field strengths. When the E-field between the electrode plates is extremely low, the Stark shift is so minimal that it is indiscernible illustrated in the inset. The minimum detectable field strength is approximately 10 V/m. }
\label{fig:calibration curve}
\end{figure}\par
\begin{figure}[ht]
\centering
\fbox{\includegraphics[width=0.7\linewidth]{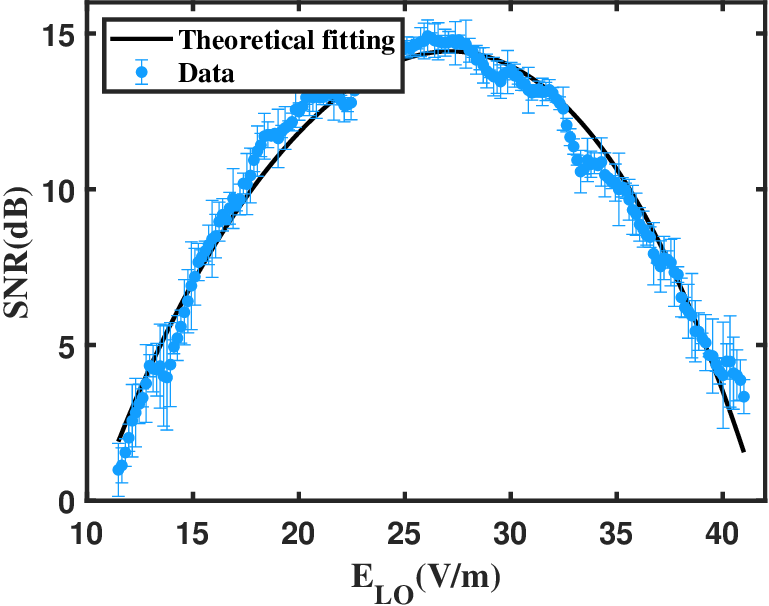}}
\caption{Optimize the LO field. SNR increase as a function of LO field. The points represent experimental data and the black line is the theoretical predictions of the model \ref{eq:sensitivity} presented in the text. The optimal value of the LO field is 26.2 V/m.}
\label{fig:optimize the local}
\end{figure}\par
In our experiments, we firstly employ 63 MHz RF E-field between the two parallel electrode plates to measure the Stark shift in the Rydberg EIT under the varying E-field strength\cite{liu2022highly,Prajapati:21}. The RF E-field with 63 MHz does not couple any Rydberg atomic resonance transitions for the Rydberg state $43D_{5/2}$. According to eq. \ref{eq:Stark shift}, the E-field-dependent frequency shift demonstrates a quadratic relationship. When the external electric field is very small, due to the limitations imposed by the EIT line width, it is no longer possible to accurately calculate the applied E-field through the Stark shift according to the movement of the EIT window illustrated in inset of fig. \ref{fig:calibration curve}. The minimum detectable E-field is approximately 10 V/m. In order to detect weaker E-field, the LO E-field and the signal field are both applied to the parallel electrode plates to accompolish heterdyne measurement\cite{liu2022highly}. The frequency of the LO field is 63.1 MHz to obtain the mediate frequency output signal with 100 kHz in which lower noise level. The strength of measured E-field is fixed, and the spectrum SNR within the LO E-field amplitude 10-40 V/m is measured. The measurement results are shown in fig. \ref{fig:optimize the local}, the SNR has a maximum value with LO field 26.2V/m. By applying our theoretical model, we investigated the response of Rydberg atoms to external fields through optical reading. Data fitting was performed, and the obtained results show excellent agreement with the dataset.
This model explains that when utilizing the heterodyne technique based on the Stark effect to read the measured E-field through EIT transmission spectra, the addition of the LO field does not indiscriminately increase the output SNR. Instead, there exists an optimal operating point.\par

\begin{figure}[ht]
\centering
\fbox{\includegraphics[width=0.7\linewidth]{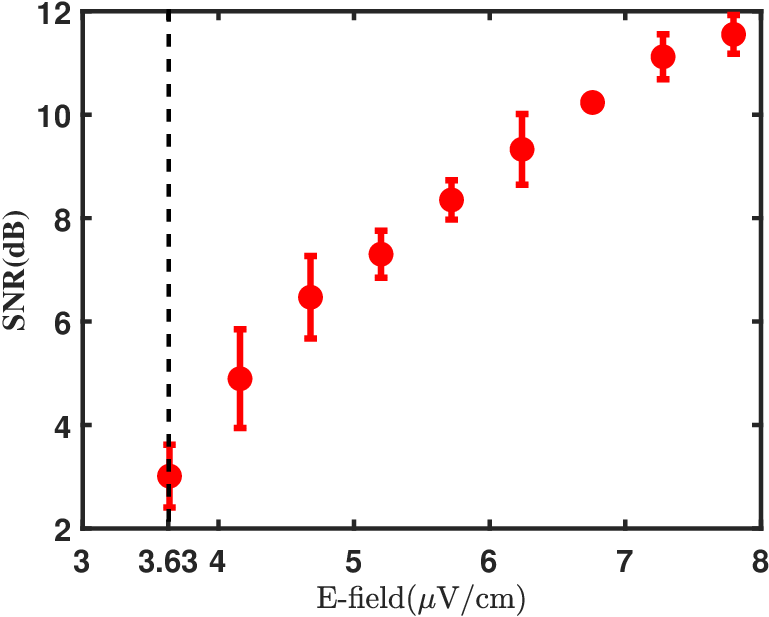}}
\caption{Plot for detecting weak E-field based on heterodyne scheme. The points represent the SNR vs. the signal E-field. The SNR indicated by the black dashed line is 3dB, corresponding to a weak field strength of 3.63 $\mu$V/cm.}
\label{fig:weak signal measurement}
\end{figure}

 The intensity of the LO field is fixed at the optimal operating point. Adjusting the value of the signal field until the SNR reaches 3 dB, corresponding to an E-field 3.63 $\mu$V/cm, as shown in fig. \ref{fig:weak signal measurement}. So the minimum detectable field strength is determined to be 2.56 $\mu$V/cm when the SNR is overwhelmed by noise.  \par
\par
\begin{figure}
    \centering
    \fbox{\includegraphics[width=\linewidth]{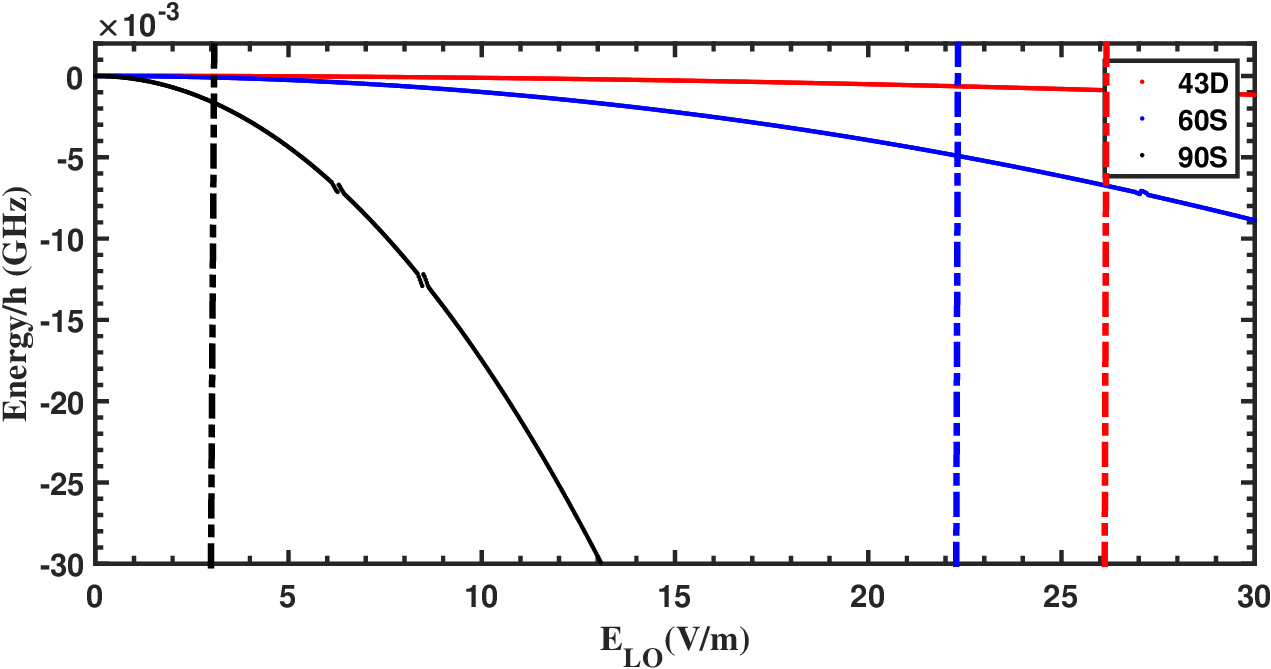}}
    \caption{Calculated Stark map of the $\rm 43D_{5/2}$, $\rm 60S_{1/2}$, and $\rm 90S_{1/2}$ states.The dashed lines represent the optimal operating points of the LO field for each of the three atomic states, with their corresponding values being 26.2V/m, 22.3V/m, and 3V/m, respectively.}
    \label{fig:stark map}
\end{figure}

We conducted sensitivity measurements on RF E-field with same frequency for the $\rm 60S_{1/2}$ and $\rm 90S_{1/2}$ atomic states. The measurement results are shown in table \ref{tab:comparison-sensitivity}. The results indicate that the sensitivity of the $\rm 90S_{1/2}$ state, which has the maximum polarization, is the best and an order of magnitude higher than the research efforts on MHz E-field during the same period as shown in the first row of table \ref{tab:comparison-sensitivity}. In order to facilitate a comparison of multiple energy levels, fig. \ref{fig:stark map} illustrates the relative frequency shifts of the $\rm 43D_{5/2}$, $\rm 60S_{1/2}$, and $\rm 90S_{1/2}$ states. The frequency shift rate of the $\rm 43D_{5/2}$ state, specifically at the optimal operating point of the LO field indicated by the red dashed line, is smaller than that of the $\rm 60S_{1/2}$ state. However, its sensitivity is better. This also demonstrates, as shown in eq. \ref{eq:Stark shift1}, that the gain depends on the combined effect of polarizability and LO field strength. Additionally, for $\rm 90S_{1/2}$ state with higher principal quantum numbers having the more dense energy level spacing than $\rm 43D_{5/2}$ and $\rm 60S_{1/2}$, the Stark curve may exhibit smearing at smaller external fields due to the mixing between nearby states, resulting in an earlier appearance of the optimal operating point for the LO field strength\cite{jau2020vapor}. Additionally, the non-uniform E-field formed on the inner surface of the vapor cell due to collisional ionization of Rydberg atoms may also contribute to the broadening of the EIT line shape\cite{jau2020vapor}. This results in the fact that although the polarizability is significant for the $\rm 90S_{1/2}$ state, the optimal operating point for the local oscillator field is relatively small.\par
\begin{figure}
    \centering
    \fbox{\includegraphics[width=0.7\linewidth]{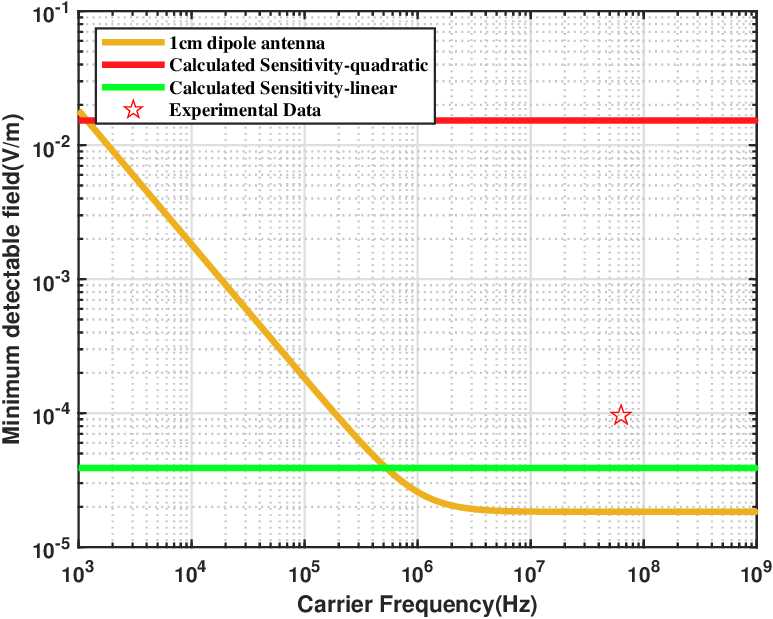}}
    \caption{Minimum detectable field as a function of carrier frequency for  theoretical calculation of the 1cm antenna systems and our experimentally measured data point using $\rm 90S_{1/2}$ state.The red(green) lines show the minimum detectable field with quadratic signal scaling(linear signal scaling)\cite{meyer2020assessment} }
    \label{fig:1 cm dipole antenna}
\end{figure}

In order to evaluate the performance of MHz E-field Rydberg sensor, fig.\ref{fig:1 cm dipole antenna} shows minimum detectable field for theoretical calculation of the 1 cm antenna systems\cite{meyer2020assessment} and our experimentally measured data point.
The solid red(green) lines show represent theoretical calculations of the minimum field for quadratic(linear) relationships. Here, we consider the ground state atomic density is $4\times 10^{11} \rm cm^{-3}$ and the density of Rydberg atoms in the vapor cell is 0.25$\%$ of the ground state atomic density. The transit time of atoms and collisions between atoms affect the coherence time, leading to the broadening of the EIT window. At this density of Rydberg atoms, the transit time relaxation rate is greater than that caused by collisions between Rydberg atoms\cite{wang2023noise}, thereby dominating the coherence time in the sensitivity calculation. According to $\Gamma_{\rm tran} =\rm \frac{1.13v}{D}$( the average velocity of the atoms v is 217 m/s,
FHWM of Gaussian laser beam D is 1 mm)\cite{sagle1996measurement}, the transit time relaxation rate is 245 kHz. This outcome at this frequency point (60MHz) approaches closely to the theoretical limit of the RF dipole antenna. Because the measurement sensitivity of the RF field based on the Stark effect of Rydberg atoms is independent of the carrier frequency, we expect to achieve sensitivity surpassing that of dipole antennas when the RF field frequency is below 1 MHz by selecting appropriate Rydberg states and optimizing the strength of LO field in case of eliminating the screening effect\cite{jau2020vapor}.

\begin{table}[htbp]
\centering
\caption{\bf Comparative study on sensitivity}
\begin{tabular}{ccccc}
\hline
\textit{nl} & $\alpha(\rm \frac{MHz\cdot cm^{2}}{V^{2}})$ & f(MHz) & $\Delta E_{min}(\frac{\mu\rm V}{\rm cm}) $ & Work \\
\hline
$\rm 55P_{3/2}$ & 2500 & 30 & 37.3 & \cite{liu2022highly}\\
$\rm 100S_{1/2}$ & 8856 & sub-kHz & 3.4 & \cite{jau2020vapor}\\
$\rm 43D_{5/2}$ & 410 & 63 & 2.56 & this work \\
$\rm 60S_{1/2}$ & 197 & 63 & 3.35 & this work\\
$\rm 90S_{1/2}$ & 3505 & 63 & 0.96 & this work\\
\hline
\end{tabular}
  \label{tab:comparison-sensitivity}
\end{table}
\section{Conclusion}
In this article, we examine and discuss the influence of polarizability and LO field strength on the sensitivity of RF E-field measurements based on Rydberg atoms in a heterodyne scheme. We quantified and provided the transmission coefficient from the E-field to be measured to the readout signal. The experimental results and theoretical model indicate that the LO field, as the gain factor of the signal field, has specific gain values for different atomic states. Our experimental results approach the theoretical sensitivity limit of RF dipole antennas and have the potential to surpass dipole antennas in the sub-MHz range. \par

\section{Backmatter}

\begin{backmatter}
\bmsection{Funding} This work was supported by the National Key R$\&$D Program of China (2022YFA1404003), the National Natural Science Foundation of China (Grant Nos. 61827824 and 61975104), the Fund for Science and Technology on Electronic Information Control Laboratory and the Fund for Shanxi ‘1331 Project’ Key Subjects Construction, Bairen Project of Shanxi Province, China.

\bmsection{Disclosures} The authors declare no conflicts of interest.
\bmsection{Data availability} Data underlying the results presented in this paper are not publicly available at this time but may be obtained from the authors upon reasonable request.
\end{backmatter}

\section{References}

\bibliography{main}

\bibliographyfullrefs{main}


\ifthenelse{\equal{\journalref}{aop}}{%
\section*{Author Biographies}
\begingroup
\setlength\intextsep{0pt}
\begin{minipage}[t][6.3cm][t]{1.0\textwidth} 
  \begin{wrapfigure}{L}{0.25\textwidth}
    \includegraphics[width=0.25\textwidth]{john_smith.eps}
  \end{wrapfigure}
  \noindent
  {\bfseries John Smith} received his BSc (Mathematics) in 2000 from The University of Maryland. His research interests include lasers and optics.
\end{minipage}
\begin{minipage}{1.0\textwidth}
  \begin{wrapfigure}{L}{0.25\textwidth}
    \includegraphics[width=0.25\textwidth]{alice_smith.eps}
  \end{wrapfigure}
  \noindent
  {\bfseries Alice Smith} also received her BSc (Mathematics) in 2000 from The University of Maryland. Her research interests also include lasers and optics.
\end{minipage}
\endgroup
}{}

\end{document}